\documentclass[review]{elsarticle}

\journal{Physics Letters B}

\usepackage{bm}
\usepackage{color}
\usepackage{ulem}
\usepackage{braket}

\begin{document}

\begin{frontmatter}

\title{Halo-induced large enhancement of soft dipole excitation of $^{11}$Li observed via proton inelastic scattering}


\author[mymainaddress,mysecondaryaddress,mythirdaddress]{J. Tanaka\corref{mycorrespondingauthor}}
\cortext[mycorrespondingauthor]{Corresponding author.}
\ead{jtanaka@ikp.tu-darmstadt.de}
\address[mymainaddress]{RCNP, Osaka University, 10-1, Mihogaoka, Ibaraki, Osaka 567-0047, Japan}
\address[mysecondaryaddress]{Department of Physics, Konan University, Higashinada, Kobe, Hyogo 658-8501, Japan}
\address[mythirdaddress]{Institute f\"ur Kernphysik, Technische Universit\"at Datmstadt, 64289 Darmstadt, Germany}
\author[3address]{R. Kanungo}
\address[3address]{Astronomy and Physics Department, Saint Mary's University, Halifax, Nova Scotia B3H 3C3, Canada}

\author[4address]{M. Alcorta}
\address[4address]{TRIUMF, Vancouver, V6T2A3, Canada}

\author[mymainaddress]{N. Aoi}%

\author[5address]{H. Bidaman}%
\address[5address]{Department of Physics, University of Guelph, Guelph N1G2W1, Canada}
\author[5address]{C. Burbadge}%
\author[4address]{G. Christian}
\author[4address]{S. Cruz}%
\author[4address]{B. Davids}%
\author[5address]{A. Diaz Varela}%
\author[4address,6address]{J. Even}%
\address[6address]{KVI-CART, University of Groningen, 9747 AA Groningen, The Netherlands}
\author[4address]{G. Hackman}%
\author[6address]{M.N. Harakeh}%
\author[4address]{J. Henderson}%
\author[7address]{S. Ishimoto}%
\address[7address]{High Energy Accelerator Research Organization (KEK), Ibaraki 305-0801, Japan}
\author[3address]{S. Kaur}%
\author[3address]{M. Keefe}%
\author[4address,8address]{R. Kr$\ddot{\rm{u}}$cken}%
\address[8address]{Department of Physics and Astronomy, University of British Columbia, Vancouver, BC V6T 1Z1, Canada}
\author[4address]{K.G. Leach}%
\author[4address]{J. Lighthall}%
\author[4address,9address]{E. Padilla Rodal}%
\address[9address]{Instituto de Ciencias Nucleares, UNAM, Mexico City 04510, Mexico}
\author[3address]{J.S. Randhawa}%
\author[4address]{P. Ruotsalainen}%
\author[3address,4address]{A. Sanetullaev}%
\author[4address]{J.K. Smith}%
\author[3address]{O. Workman}%
\author[10address,mymainaddress]{I. Tanihata}%
\address[10address]{School of Physics and Nuclear Energy Engineering and IRCNPC, Beihang University, Beijing 100191, China}


\begin{abstract}
Proton inelastic scattering off a neutron halo nucleus, $^{11}$Li, has been studied in inverse kinematics at the IRIS facility at TRIUMF. The aim was to establish a soft dipole resonance and to obtain its dipole strength. Using a high quality 66 MeV $^{11}$Li beam, a strongly populated excited state in $^{11}$Li was observed at $E_x$=0.80 $\pm$ 0.02 MeV with a width of $\Gamma=$ 1.15 $\pm$ 0.06 MeV. A DWBA (distorted-wave Born approximation) analysis of the measured differential cross section with isoscalar macroscopic form factors leads to conclude that this observed state is excited in an electric dipole (E1) transition. Under the assumption of isoscalar E1 transition, the strength is evaluated to be extremely large amounting to 600$\sim$2000 Weisskopf units, exhausting 4\%$\sim$14\% of the isoscalar E1 energy-weighted sum rule (EWSR) value. The large observed strength originates from the halo and is consistent with the simple di-neutron model of $^{11}$Li halo.
\end{abstract}


\end{frontmatter}

Understanding the $^{11}$Li structure is a landmark in studies of the halo nuclei \cite{Tanihata19852676,Hansen1987409}. The two valence neutrons in $^{11}$Li have a very low separation energy, forming a low-density halo. As a collective excitation of the two-neutron halo, the soft-dipole resonances in $^{11}$Li are expected to appear at low excitation energies \cite{Ikeda1992355} and by nature of the excitation should have both isovector and isoscalar components. The soft-dipole resonances in $^{11}$Li have been predicted to appear at around 0.7 MeV and 2.7 MeV by the cluster-orbital shell model (COSM) that is constructed in a microscopic framework as the three-body system $^9$Li + $n$ + $n$. These low-lying states are predicted to exhaust 8\% of the isovector E1 EWSR \cite{Suzuki1990599}.\par
Several Coulomb-dissociation experiments have been performed at relatively high bombarding energies on Pb targets to reveal the E1 strength at low excitation energy.
In the early 1990's, experiments at MSU \cite{Ieki1993730} at 24 MeV/u and RIKEN \cite{Shimoura199529} at 64 MeV/u reported the experimental results showing a strong dipole strength at low excitation energy.
Later, measurements at GSI \cite{Zinser1997151} at 280 MeV/u reported that there were dipole states centered at 1.0 $\pm$ 0.1 MeV and 2.4 $\pm$ 0.1 MeV in $^{11}$Li, and that these two states exhausted 8\% of the isovector E1 EWSR, in good agreement with the COSM prediction.
The experiment at RIKEN \cite{Nakamura2006252502}, on the other hand, indicated a peak at lower energy $\sim$ 0.6 MeV that was interpreted as a soft-dipole excitation. The isovector $B$(E1) value integrated over E$_{\rm{rel}} <$ 3 MeV was obtained to be 1.42(18) e$^2$fm$^2$, which was the largest E1 strength observed so far for a low-lying dipole state. However, since the direct breakup mechanism is dominant for such Coulomb dissociation measurements, as discussed for $^{11}$Be \cite{Nakamura1994296}, it was considered that this low-energy E1 peak reported in Ref. \cite{Nakamura2006252502} corresponds to a direct breakup to the continuum.
Though the prediction of COSM model is a resonant excited state, the enhancement observed in Coulomb dissociation arises from the small separation energy in $^{11}$Li as predicted in Ref. \cite{Hansen1987409}.\par
\begin{figure*}[h!]
\begin{center}
\includegraphics[width=\linewidth]{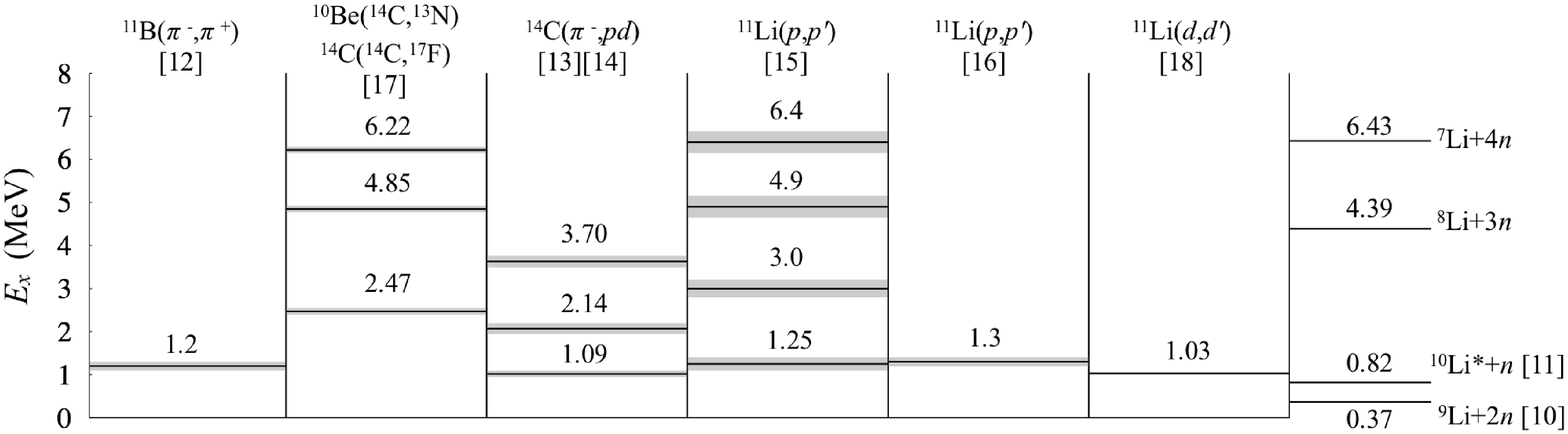}\vspace{-0.4cm}
\caption{\label{fig:ex2}The reported level schemes of  $^{11}$Li and particle decay thresholds\cite{Smith2008202501}\cite{Cavallaro2017012701}. The grey hatched bars are uncertainties in excitation energy. The references for the reaction data are denoted.}
\vspace{-0.3cm}
\end{center}
\end{figure*}

On the other hand, in the missing-mass method nuclear excitations are observed at backward scattering angles, where the effects due to the Coulomb dissociation process are negligible.  Therefore, a resonant state, if it exists, should be observed more clearly in the missing-mass spectrum due to the absence of a large E1 breakup effect. The excitation energies for $^{11}$Li obtained from several reactions are summarized in Fig. \ref{fig:ex2}. 

A pion-induced double-charge-exchange $^{11}$B($\pi^{-}$,$\pi^{+}$) reaction \cite{Kobayashi1992343}, a pion-capture reaction $^{14}$C($\pi^{-}$,$pd$) \cite{Gornov19984325,Kelley201288} and $^{11}$Li($p$,$p$$^{\prime}$) \cite{Korsheninnikov1996R537,Korsheninnikov19972317} experiments reported an excited state at around 1 MeV. However, the $^{10}$Be($^{14}$C,$^{13}$N) and $^{14}$C($^{14}$C,$^{17}$F) experiments showed that there is a state at an excitation energy of 2.47 MeV \cite{Bohlen19957}. Some of these experiments were performed with very poor resolution, and some had low statistics. Reliable information on the width and the transition strength has not been obtained so far. In order to study the resonant structure and its strength, high-statistics data with good resolution are therefore required.\par
The recent $^{11}$Li($d$,$d$$^\prime$) experiment showed a clear peak structure at around 1 MeV excitation energy \cite{Kanungo2015192502}. 
The angular distribution indicated that the excitation is due to an isoscalar E1 transition.\par The leading order operator for the isoscalar-dipole excitation is the operator $\frac{e}{2}r^3Y_1$ \cite{Harakeh2001}. In stable nuclei, the strength connected with this operator is relatively small compared with these of other types of multipole excitation partly because the $\frac{e}{2}r^3Y_1$ matrix element is suppressed by the rapid fall off of the density at large radial distances. However, since a halo nucleus has a long density-distribution tail, a strong isoscalar E1 transition strength is expected to come from a combined effect of the large $r$ (radius) and the $r^3$ factor in the operator. This strong transition strength would be a good indication that the low-lying state results from the excitation of the halo. Since a similar low-lying dipole resonance is not observed in $^9$Li, then the two-neutron halo in $^{11}$Li is the origin of this resonance.
Proton inelastic scattering at low incident energy is the simplest reaction for extracting the dipole transition strength. In addition, ($p$,$p$$^{\prime}$) is a complementary reaction to ($d$,$d$$^{\prime}$) that is necessary to establish the dipole resonance. For solving the long-standing controversy regarding the low-lying dipole resonances in $^{11}$Li, we performed a high statistics and high resolution $^{11}$Li($p$,$p$$^\prime$) measurement to determine the strength of the soft-dipole resonance. \par
\begin{figure}[h!]
\includegraphics[width=\linewidth]{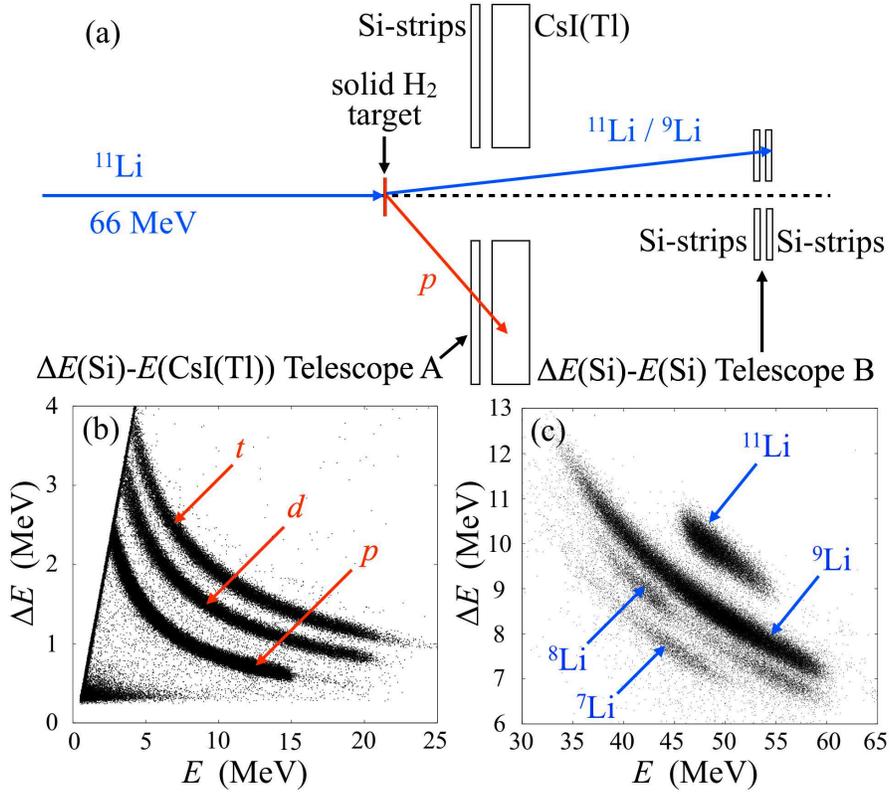}\vspace{-0.2cm}
\caption{\label{fig:setup} (a) The experimental setup at IRIS. $\Delta E-E$ particle identification spectra for (b) hydrogen isotopes and (c) Li isotopes.}
\vspace{-0.4cm}
\end{figure}

The experiment was performed at the IRIS facility at TRIUMF in Canada.
A high-quality $^{11}$Li beam at 6 MeV/u from the ISAC II facility was incident on a solid hydrogen target with a thickness of $\sim$150 $\mu$m. The target is formed on a 5 $\mu$m Ag foil backing with the foil facing the incoming beam. Therefore, the scattered protons from the H$_2$ target reach the detectors unhindered. Using a $\Delta E-E$ detector system consisting of a Si-strip detector array and CsI(Tl) detectors, an excitation energy resolution of 170 keV ($\sigma$) was achieved under a low-background condition. Figure \ref{fig:setup} shows a schematic drawing of the experimental setup and the measured spectra for particle identification of hydrogen and Li isotopes. The recoil protons were detected by Telescope A, consisting of 100 $\mu$m thick annular Si-strip detectors \cite{Davinson2000350} and annular-CsI(Tl) detectors. 
Inelastically scattered $^{11}$Li*(excited state of $^{11}$Li) decays into $^{9}$Li and two neutrons. $^{9}$Li ions are detected by Telescope B, consisting of two layers of 60 $\mu$m thick and 500 $\mu$m thick annular Si-strip detectors.

By using the energies and polar angle information of the recoil protons, the $^{11}$Li missing-mass spectrum was obtained. The coincident measurement with $^9$Li, emitted after the $^{11}$Li decay, improves the selection of the $^{11}$Li($p$,$p$$^{\prime}$) reaction channel. Moreover, a coplanarity gate on the relative azimuthal angle between proton and $^9$Li, which was defined by $\phi_{^{11}\rm{Li}-p}$=180$^\circ$$\pm$ 22.5$^\circ$, decreased the background from the non-resonant decay events. After a two-body reaction of inelastic scattering, the $^9$Li decay residue from the excited state of $^{11}$Li is emitted in almost the same direction as the excited $^{11}$Li nucleus because the decay energy is much smaller than the mass of $^9$Li. On the other hand, the $^9$Li from the direct breakup of $^{11}$Li due to interaction with the proton target will in general not necessarily be emitted in the same direction as $^{11}$Li, which is determined by the four-body final state phase space.

\begin{figure*}[h!]
\includegraphics[width=\linewidth]{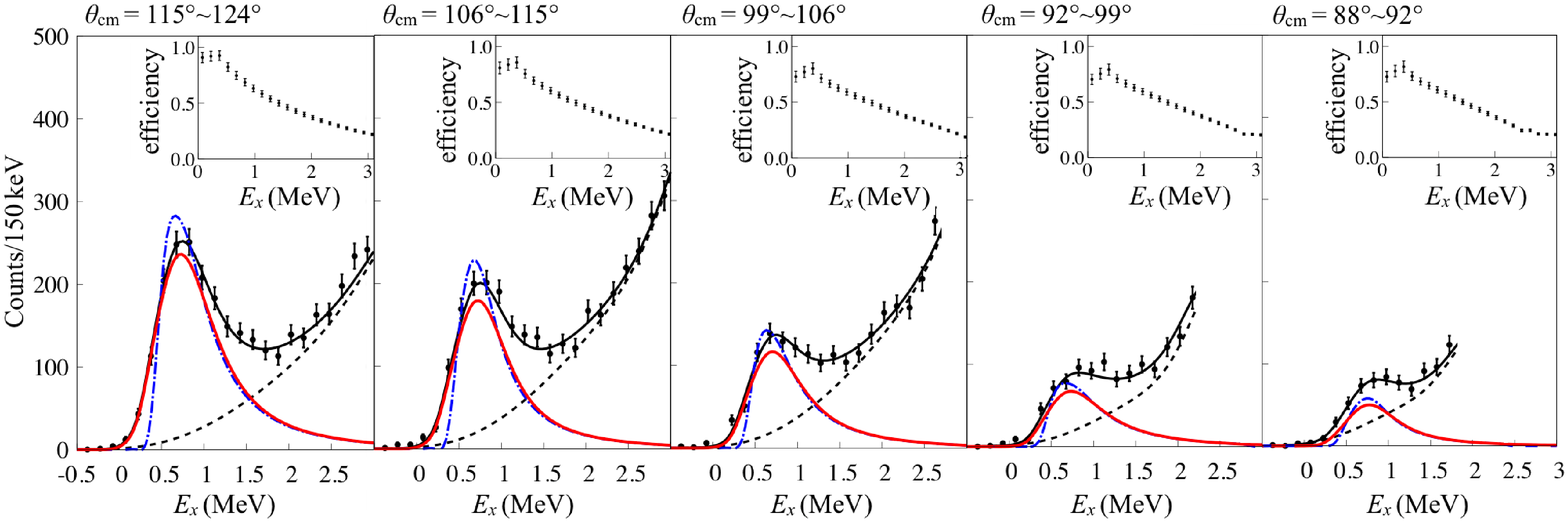}\vspace{-0.4cm}
\caption{\label{fig:inelastic_spectrum1}Excitation-energy spectra of $^{11}$Li($p$,$p$$^{\prime}$) in coincidence with $^{9}$Li at different scattering angles. The red solid lines show the fitted spectrum with a Breit-Wigner function folded with the excitation-energy resolution of the detector system. The original Breit-Wigner function is shown by the blue dash-dotted lines and starts at the two-neutron break-up threshold. The Breit-Wigner functional forms have been corrected for the detection efficiency as function of excitation energy as shown in the insets. The dotted lines are the results of the four-body phase-space calculations, expected as continuum spectra. The solid black curves show the total fits, i.e. sums of red solid curves and blue dash-dotted curves.}\vspace{-0.4cm}
\end{figure*}\par

The obtained excitation-energy spectra with their energy-dependent detection efficiencies are shown in Fig. \ref{fig:inelastic_spectrum1}. These efficiencies were calculated by taking into account both the coincidence-gate efficiency and the coplanarity-gate efficiency resulting from the detector geometry and the angular spread from the decay of $^{11}$Li to $^{9}$Li.
The $^{11}$Li($p$,$p^{\prime}$) spectra at the different angles were fitted to obtain the resonant energy, the width and the differential cross sections.
A Breit-Wigner function $F(E_r)$ with an energy-dependent width $\Gamma(E_r)$ was employed to fit the spectra assuming a resonant state near the particle decay threshold. 
The function $F(E_r)$ is expressed as,
\vspace*{-0.1cm}
 \begin{equation}
F(E_r)=\frac{\Gamma(E_r)}{(E_{x}-E_{0})^2+\Gamma^2(E_r)/4},
 \vspace*{-0.1cm}
 \end{equation}
where $E_r$ is the relative energy of decay particles, $E_{0}$ is the excitation energy of the resonant peak observed in $^{11}$Li. The relationship between these variables is $E_{x}$=$E_{s}$+$E_{r}$, where $E_{s}$ is the 2n separation energy.
The width $\Gamma(E_r)$ is a function of energy defined as $\Gamma(E_r)\equiv g\sqrt{E_r}$, where $g$ is a fitting parameter.
The experimental energy resolution was taken into account by folding the Breit-Wigner function with a Gaussian of $\sigma$=170 keV, which was obtained from fitting the elastic scattering peak in the $^{11}$Li excitation-energy spectrum.
The peak position and the resonant width were determined consistently by fitting all the spectra at the different scattering angles to be $E_{0}$=0.80 $\pm$ 0.02 MeV and $\Gamma(E_{r})=1.15$ $\pm$ 0.06 MeV.\par
\begin{figure}[h!]
\includegraphics[width=\linewidth]{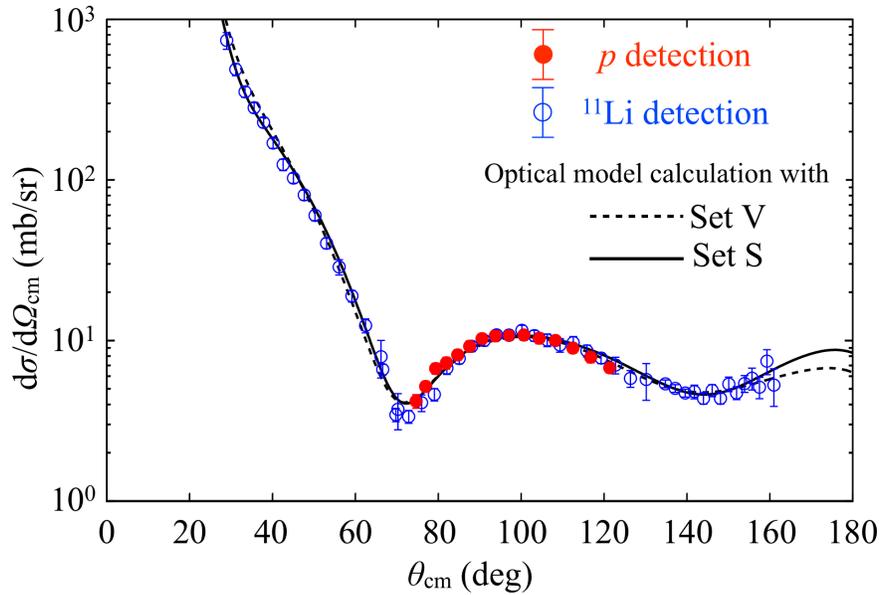}
\caption{\label{fig:cross_all}The differential cross sections of the $^{11}$Li+$p$ elastic scattering at 66 MeV. The blue open circles are the experimental data from $^{11}$Li detection. The red filled circles are from proton detection. Fitted results with Set S and Set V are shown by the black solid line and dotted line, respectively. The experimental angular resolution was included in the calculations.}
\end{figure}

Differential cross sections of the elastic scattering obtained from detection of either proton or $^{11}$Li are plotted in Fig. \ref{fig:cross_all}.
In addition to the statistical uncertainties of the data, the total systematic uncertainties were estimated to be $\pm$7\%. The contributions to the systematic uncertainties consist of 4.8\% coming from the target thickness and 5.0\% coming from the absolute counting of the incident beam.\\
 The optical potentials were obtained from the proton elastic scattering data assuming the following form:
 \begin{equation}\hspace{-2.5cm}
     U(r) =-V_vf(r,r_v,a_v)+4\left(\frac{\hbar}{m_{\pi}c}\right)^{2}\frac{1}{r}\frac{d}{dr}\{V_{so}f(r,r_{so},a_{so})\}\bm{l}\cdot\bm{s}+V_C(r_C)+i4a_s\frac{d}{dr}\{W_sf(r,r_s,a_s)\}- iW_wf(r,r_{w},a_{w})\\
\end{equation}
where the Woods-Saxon potential shape $f(r,r_i,a_i)=\left\{1+e^{(r-r_iA^{1/3})/a_i}\right\}^{-1}$ was used.
\par
The obtained optical potential parameter sets with the imaginary part having only the volume term (Set V) and with only the surface imaginary term (Set S) are listed in Table \ref{tab:table3}.
\begin{table*}[h!]
\caption{\label{tab:table3} The optical potential parameters Set V and Set S obtained by fitting of the elastic scattering data.}
\hspace{-3.0cm}
\begin{tabular}{ccccccccccccccc}
\hline\hline
&V$_v$&r$_v$&a$_v$&V$_{so}$&r$_{so}$
&a$_{so}$&r$_C$&W$_w$&r$_w$&a$_w$&W$_s$&r$_s$&a$_s$&$\chi_{red.}^2$\\
&(MeV)&(fm)&(fm)&(MeV)&(fm)&(fm)&(fm)&(MeV)&(fm)&(fm)&(MeV)&(fm)&(fm)&\\ \hline
Set V&54.2&1.16&0.75&6.23&1.16&0.75&1.16&14.3&0.61&1.98&0.0&-&-&1.81\\
Set S&35.2&1.71&0.92&7.95&1.71&0.92&1.71&0.0&-&-&14.4&1.49&0.54&1.05\\
\hline\hline
\end{tabular}
\end{table*}
\par
\begin{figure}[h!]
\includegraphics[width=\linewidth]{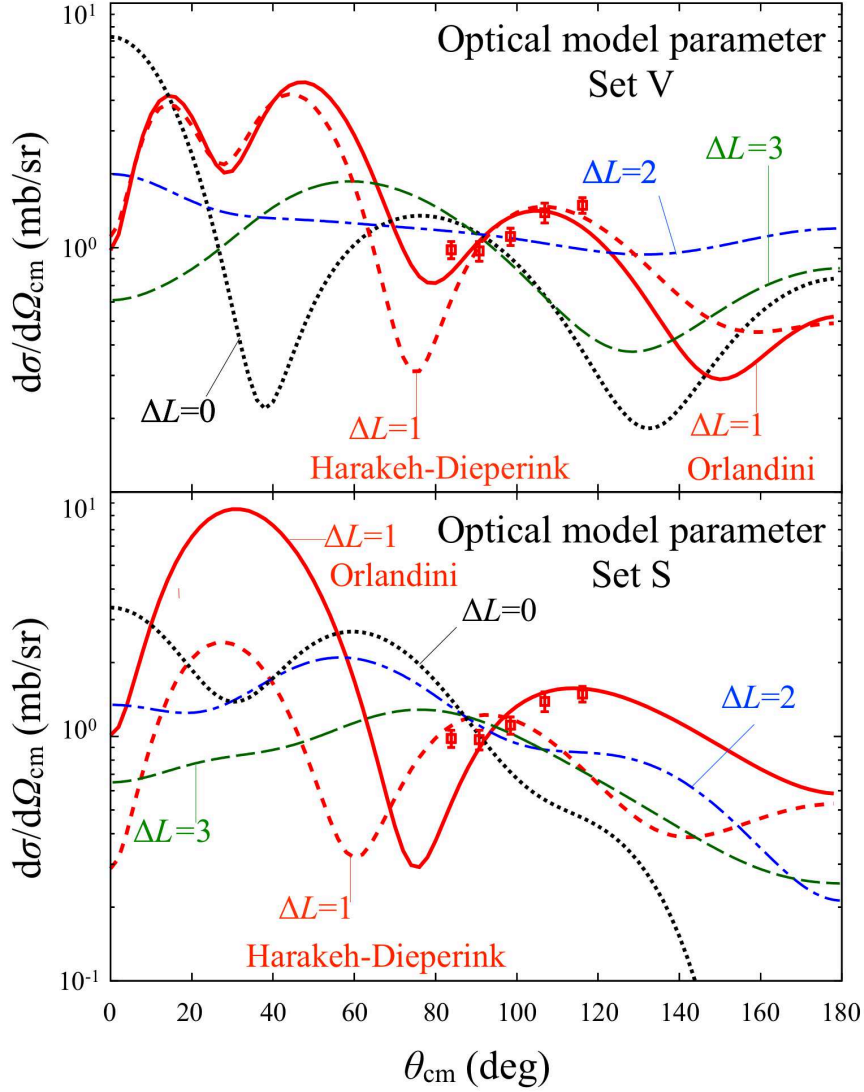}\vspace{-0.0cm}
\caption{\label{fig:DWBA} The measured differential cross sections of inelastic scattering in comparison with the DWBA calculations with optical potential Set V (top) and Set S (bottom). The black dotted lines are for $\Delta L$=0, the red dashed lines are for $\Delta L$=1 with the Harakeh-Dieperink form factor. The red solid lines are for $\Delta L$=1 with the Orlandini form factor. The blue dash-dotted lines are for $\Delta L$=2. The green long-dashed lines are for $\Delta L$=3.}
\end{figure}

The inelastic-scattering differential cross sections (empty red squares with error bars) were compared to DWBA predictions using the code, CHUCK3 \cite{Kunz1977}, shown in Fig \ref{fig:DWBA}. 
Different form factors were used for different multipolarities of transition $\Delta L$ between the ground state and the observed excited state. The optical potential for the exit channel was assumed to be the same for the entrance channel. For the very low bombarding energy and the low-Z of the hydrogen target, Coulomb excitation of isovector dipole strength, in particular, is expected to be negligible at the backward center-of-mass angle at which measurements were made in this experiment. Furthermore, the isoscalar excitation was expected to be dominant in case of the present low-energy $^{11}$Li($p$,$p$$^\prime$) experiment \cite{Love19811073}. For $\Delta L$=2 (quadrupole) and $\Delta L$=3 (octupole), the form factors obtained in the surface vibrational model \cite{Bohr1975} were used. In such models, the nuclear shape vibrates according to quadrupole or octupole deformations without changing the density. For $\Delta L$=0, the breathing-mode form factor was used \cite{Satchler1987215}. It changes the nuclear size and the density changes by conserving the number of nucleons. For $\Delta L$=1, the Harakeh-Dieperink form factor \cite{Harakeh19812329} and the Orlandini form factor \cite{Orlandini198221} were used. These form factors for $\Delta L$=1 were introduced to describe the isoscalar E1 excitations with the $\frac{e}{2}r^3Y_1$ operator.
The Harakeh-Dieperink form factor is obtained from a sum-rule approach model (doorway dominance model) and is most appropriate for the collective 3$\hbar\omega$ compression mode exhausting the largest fraction of the isoscalar dipole EWSR. Instead, the Orlandini form factor is determined to describe 1$\hbar\omega$ excitations with a very small fraction of the isoscalar E1 EWSR.

\vspace{-0.0cm}

The DWBA calculations were performed for both optical parameter sets V and S. The results of the calculations are summarized in Table \ref{strength}. The absolute value was normalized to fit the experimental data.
The $\Delta L$ = 0, 2, 3 angular distributions show negative slopes at $\theta_{CM}$ $\sim$ 90$^\circ$, which are different from the data. Only the $\Delta L$=1 calculations show distributions that are in closest agreement with the data. The calculation using the volume imaginary potential provides the best fit to the data, both using the Harakeh-Dieperink (H-D) and Orlandini (O) form factors.\par
The transition strength was evaluated from the best-fit amplitude of the DWBA calculations. 
The transition strength to the 0.80 MeV state was extremely large amounting to 600 $\sim$ 2000 Weisskopf units (W.u.) and 4\% $\sim$ 14\% of the isoscalar E1 EWSR.
\begin{center}
\begin{table}[h!]
\caption{\label{strength}Strengths of the isoscalar E1 excitation $B$(E1(IS)) following the prescription written in \cite{Harakeh19812329} ; comparison of the strength with the Weisskopf unit $B_{\rm{W}}$(E1(IS)) and the EWSR value $S$(E1(IS)). The last row is the strength estimated with the assumption of a di-neutron wave function in the square-well potential model.}
\begin{center}
\begin{tabular}{ccccc}\hline\hline
Optical&Form&$B$(E1(IS))&$\frac{B(\rm{E1(IS)})}{B_{\rm{W}}(\rm{E1(IS)})}$
&$S$(E1(IS))\\
potential&Factor&($\times10^3 $e$^2$fm$^6$)&($\times10^2$ W.u.)&(\%)\\ \hline
 Set V&H-D&2.6&14&9.4 $\pm$ 1.9\\
 Set V&O&1.1&6.1&3.9 $\pm$ 2.2\\
 Set S&H-D&3.6&20&13.4 $\pm$ 2.2\\
 Set S&O&3.8&21&13.9 $\pm$ 2.2\\\hline
 Di-neutron&model&0.5&2.8&-\\\hline\hline
\end{tabular}
 \end{center}
\end{table}
\end{center}
This large strength can be qualitatively understood as the feature of the isoscalar E1 operator $\frac{e}{2}r^3Y_1$ together with the spatially extended neutron-halo structure.
The strength enhanced by this effect is well estimated by introducing a di-neutron weakly bound in the square-well potential. Using the extended halo distribution, it was found that the transition strength is $\sim$ 2.8 W.u., which is of the similar order of magnitude as the present experimental result. In the actual calculation, the operator $\frac{e}{2}\left\{r^3-\frac{5}{3}\braket{r^2}r\right\}Y_1$, corrected for the center-of-mass motion, was used to calculate the transition rate.
This simple model estimation shows that the strength comes from the low-density halo far outside of the square-well potential.\par
The observed peak in the $^{11}$Li($p$,$p$$^{\prime}$) spectrum is slightly lower in energy and a bit wider than the one found in the $^{11}$Li($d$,$d$$^\prime$) experiment. Dipole transitions from the 3/2$^{-}$ ground state of $^{11}$Li can lead to states with spins of 1/2$^+$, 3/2$^+$ and 5/2$^+$. If two or more of these states are populated through this soft dipole excitation in the ($p$,$p$$^\prime$) reaction, and if they are relatively closely spaced, the observed state in ($p$,$p$$^\prime$) experiment could include two unresolved dipole states. This may account for the small difference in peak position seen compared to ($d$,$d$$^\prime$).\par
In summary, a low-energy dipole excitation state at 0.80 MeV in $^{11}$Li has been identified in low-energy proton inelastic scattering off $^{11}$Li. The measured angular distribution of the differential cross sections is consistent with predictions using the form factor for an isoscalar E1 excitation mode. A very large isoscalar E1 transition probability of 1.1 $\sim$ 3.8$\times 10^{3}$ e$^2$fm$^6$ is deduced, exhausting 4 $\sim$ 14\% of the isoscalar E1 EWSR. This large dipole strength is found to originate from the halo and is consistent with a simple di-neutron model for $^{11}$Li. The results bring new information on the soft dipole excitation. The Current derivation of E1 strength assumes that the observed peak is described only by an isoscalar excitation. While it is true that the (p,p') reaction can excite both isoscalar and isovector modes, as described above, the isoscalar excitation may by expected to be dominant here. However future theoretical developments considering both isoscalar and isovector descriptions for the soft-dipole excitation should lead to a deeper understanding, which is beyond the scope of this article.\par
%
%
%
The experiment was partly supported by the grant-in-aid program of the Japanese government under the contract number 23224008 and 14J03935.
The work is supported by NSERC, Canada foundation for innovation and Nova Scotia Research and Innovation Trust. TRIUMF receives funding via a contribution through the National Research Council Canada.
The support of the PR  China government and Beihang University under the Thousand Talent program is gratefully acknowledged.
J.T. gratefully acknowledges the support by the Hirao Taro Foundation of the Konan University Association for Academic Research.
Discussions with Dr. K. Ogata and Dr. T. Matsumoto are gratefully acknowledged. 
%

\section*{References}
\begin{thebibliography}{9}
%
%
%
%
\bibitem{Tanihata19852676}I. Tanihata $et$ $al.$, Phys. Rev. Lett. $\bm{55}$, 2676 (1985).
\bibitem{Hansen1987409}P. G. Hansen and B. Jonson, Europhys. Lett. $\bf{4}$, 409 (1987).
\bibitem{Ikeda1992355}K. Ikeda, Nucl. Phys. $\bf{A538}$, 355c (1992).
\bibitem{Suzuki1990599}Y. Suzuki $et$ $al.$, Nucl. Phys. $\bf{A517}$, 599 (1990).
%
%
\bibitem{Ieki1993730}K. Ieki $et$ $al.$, Phys. Rev. Lett. $\bm{70}$, 730 (1993).
\bibitem{Shimoura199529}S. Shimoura $et$ $al.$, Phys. Lett. B $\bm{348}$, 29 (1995).
\bibitem{Zinser1997151}M. Zinser $et$ $al.$, Nucl. Phys. $\bf{A619}$, 151 (1997).

\bibitem{Nakamura2006252502}T. Nakamura $et$ $al.$, Phys. Rev. Lett. $\bm{96}$, 252502 (2006).
\bibitem{Nakamura1994296}T. Nakamura $et$ $al.$, Phys. Lett. B $\bm{331}$, 296 (1994).
  \bibitem{Smith2008202501}  M. Smith $et$ $al.$, Phys. Rev. Lett. $\bf{101}$, 202501 (2008).
    \bibitem{Cavallaro2017012701}M. Cavallaro $et$ $al.$, Phys. Rev. Lett. $\bm{118}$, 012701 (2017).
  \bibitem{Kobayashi1992343}T. Kobayashi, Nucl. Phys. $\bf{A538}$, 343c (1992).
  \bibitem{Gornov19984325}M. G. Gornov $et$ $al.$, Phys. Rev. Lett. $\bf{81}$, 4325 (1998).
  \bibitem{Kelley201288}J.H. Kelley $et$ $al.$, Nucl. Phys. $\bf{A880}$, 88 (2012).
\bibitem{Korsheninnikov1996R537} A. A. Korsheninnikov $et$ $al.$, Phys. Rev. C $\bm{53}$, R537 (1996).
\bibitem{Korsheninnikov19972317} A. A. Korsheninnikov $et$ $al.$, Phys. Rev. Lett. $\bm{78}$, 2317 (1997).
\bibitem{Bohlen19957}H. G. Bohlen $et$ $al.$, Z. Phys.  A $\bf{351}$, 7 (1995).
  \bibitem{Kanungo2015192502}R. Kanungo $et$ $al.$, Phys. Rev. Lett. $\bm{114}$, 192502 (2015).
  \bibitem{Harakeh2001}M. N. Harakeh and A. van der Woude, $Giant$ $Resonances$, Oxford University Press (2001).
\bibitem{Davinson2000350}T. Davinson $et$ $al.$, Nucl. Instr. Meth. Phys. Res. A $\bf{454}$, 350 (2000).
\bibitem{Kunz1977}P. D. Kunz, CHUCK- A Coupled-Channel Code, University of Colorado (1977) unpublished; modified by J.R. Comfort and M.N. Harakeh.
\bibitem{Love19811073} W. G. Love $et$ $al.$, Phys. Rev. C 24, 1073 (1981).
\bibitem{Bohr1975}A. Bohr and B. R. Mottelson, $Nuclear\,Structure\,$I\hspace{-.1em}I, World Scientific Publishing Co. (1975).
\bibitem{Satchler1987215}G. R. Satchler, Nucl. Phys. $\bf{A472}$, 215 (1987).
\bibitem{Harakeh19812329}M. N. Harakeh and A. E. L. Dieperink, Phys. Rev. C $\bf{23}$, 2329 (1981).
\bibitem{Orlandini198221}G. Orlandini $et$ $al.$, Phys. Lett. B $\bf{119}$, 21 (1982).
\end{thebibliography}
\end{document}